\documentclass[
  english,
  prb,
  amsmath,
  amssymb,
  reprint,
  nofootinbib,
  onecolumn
  ]{revtex4-2}

\usepackage[utf8]{inputenc}
\usepackage{babel}
\usepackage{graphicx}
\usepackage[unicode=true,pdfusetitle,
  bookmarks=true,bookmarksnumbered=false,bookmarksopen=false,
  breaklinks=false,pdfborder={0 0 1},backref=true,colorlinks=true]
 {hyperref}
\hypersetup{pdfborderstyle=}

\usepackage{dsfont}
\usepackage{braket}
\DeclareMathOperator{\Tr}{Tr}
\usepackage{bm}
\usepackage{orcidlink}

\begin{document}

\title{Extended Landau--Lifshitz equation for nanomagnets: a path-integral derivation of surface-induced magnetization nutation}

\author{Hamid Kachkachi\orcidlink{0000-0002-6954-8117}}
\affiliation{Université de Perpignan Via Domitia, Laboratoire PROMES-CNRS (UPR 8521), Rambla de la Thermodynamique, Tecnosud, 66100 Perpignan, France}
\email[Corresponding author:]{hamid.kachkachi@univ-perp.fr}

\author{Pascal Thibaudeau\orcidlink{0000-0002-0374-5038}}
\affiliation{CEA, DAM, Le Ripault, F-37260, Monts, France}
\email{pascal.thibaudeau@cea.fr}
\date{\today}

\begin{abstract}
  An effective dynamical equation for the magnetization of a nanomagnet with surface anisotropy is derived from an atomistic spin Hamiltonian using the spin coherent-state path integral formalism. The derivation proceeds in two steps. First, the continuum Euclidean action for the many-spin nanomagnet is obtained, including the Wess--Zumino--Witten (Berry phase) term, as well as exchange, Zeeman, and core/surface anisotropy contributions. Second, the local magnetization density is decomposed into a slowly varying macrospin component and transverse spin-misalignment fluctuations driven by surface effects.
  A systematic expansion of the action is then performed up to quadratic order in the transverse-fluctuation variables. Under the adiabatic approximation, where transverse modes relax much faster than the macrospin, these modes are eliminated using their static Green's-function solution. This results in a closed, extended Landau--Lifshitz equation for the macrospin, featuring an effective field with nontrivial corrections from spin misalignment. These corrections renormalize both the Zeeman and anisotropy fields and introduce additional terms that act as nutation- and damping-like contributions.
  For a spherical nanoparticle with Néel surface anisotropy, the leading surface-induced correction is explicitly evaluated, and its structure is compared to inertial extensions of the Landau--Lifshitz--Gilbert equation.
  Together, these results establish a microscopic foundation for surface-induced magnetization nutation in nanomagnets and provide a framework to estimate corrections to the precession frequency and effective damping.
  The corresponding shift in the ferromagnetic-resonance frequency and linewidth is measurable with standard GHz spectrometers, and the underlying adiabatic-elimination mechanism is expected to generalize to any slow magnetic variable coupled to a bath of fast fluctuating modes.
\end{abstract}

\maketitle

\section{Introduction}
\label{sec:introduction}

The quest for ultrafast control of magnetization in nanoscale systems is a central theme in spintronics. 
Conventional magnetic switching, for instance in magnetic recording, typically relies on reversing the magnetization with an applied field opposite to the initial state, and the efficiency of this process is limited by the macroscopic relaxation time of the magnetization, of order $100\,$ps.
Faster switching schemes exploit picosecond or sub-picosecond magnetic field pulses applied perpendicular to the equilibrium magnetization direction~\citep{gerritsLargeangleMagnetizationDynamics2006,barrosOptimalSwitchingNanomagnet2011}. 
Conventionally, the dynamics of the magnetization is described as a precessional motion around an effective field, supplemented by a Gilbert damping term.

However, a misalignment between the precession axis and the angular momentum induces an additional mode of motion: \emph{magnetization nutation}.
Analogous to the nutation of a mechanical gyroscope, this motion was historically considered a minor correction to precession. 
Recently, however, it has attracted significant interest, spurred by experimental observations of THz-frequency nutational oscillations in thin films~\citep{liInertialTermsMagnetization2015,*neerajInertialSpinDynamics2021,*unikandanunniInertialSpinDynamics2022,*deMagneticNutationTransient2025}.

Nutation-like phenomena have previously been studied in a variety of contexts, including nuclear magnetic resonance~\citep{torreyTransientNutationsNuclear1949}, optical resonance~\citep{hockerObservationOpticalTransient1968}, and electron paramagnetic resonance~\citep{vermaTimeResolvedESR1973,atkinsTransientNutationsElectron1974,fedorukTransientNutationEPR2002}. 
Theoretical predictions of spin nutation have also been made in Josephson junctions~\citep{zhuElectricFieldControl2006,franssonSpinDynamicsTunnel2008,franssonDetectionSpinReversal2008,nussinovSpinSpinwaveDynamics2005,zhuNovelSpinDynamics2004}, and more recently within approaches based on relativistic quantum mechanics, first-principles calculations~\citep{mondalRelativisticTheoryMagnetic2017,mondalDynamicsRelativisticElectron2020}, and electronic structure theory~\citep{bhattacharjeeAtomisticSpinDynamic2012,fahnleGeneralizedGilbertEquation2011,kikuchiSpinDynamicsInertia2015,thonigMagneticMomentInertia2017,chengSpinmechanicalInertiaAntiferromagnets2017}. 
Another route to nutation in metallic ferromagnets proceeds via magnetic inertia, arising from the nonadiabatic response of environmental degrees of freedom~\citep{kikuchiSpinDynamicsInertia2015,makhfudzNutationWavePlatform2020,thibaudeauEmergingMagneticNutation2021}. 
At the phenomenological level, magnetic inertia has been encoded in extended versions of the Landau--Lifshitz--Gilbert equation~\citep{ciorneiMagnetizationDynamicsInertial2011,oliveFerromagneticResonanceInertial2012,oliveDeviationLandauLifshitzGilbertEquation2015,wegroweMagneticMonopoleSeparation2016} building on the classical works of Landau and Lifshitz~\citep{landauTheoryDispersionMagnetic1935} and Gilbert~\citep{gilbertFormulationFoundationsApplications1956}.
Nutation-like excitations have also been reported in systems of coupled vortices~\citep{leeNutationlikemodeExcitationCoupled2017,gareevaNutationExcitationsGyrotropic2023} and elongated nano-elements~\citep{urzagastiLocalizedPrecessionalNutational2020}.

More recently, it has been shown that spin misalignment induced by surface effects in nanoscale magnetic particles can trigger nutational oscillations of the magnetization at frequencies in the GHz--THz range~\cite{bastardisMagnetizationNutationInduced2018, adamsLowfrequencySignatureMagnetization2024, kachkachiMagnetizationNutationMagnetic2025}.
The underlying physical picture is that surface anisotropy or other lattice inhomogeneities induce a nonuniform spin texture, so that the net magnetization (macrospin) is coupled to fast transverse modes. 
Eliminating these modes leads to a renormalized effective field and to dynamical corrections that may be interpreted as nutation or inertia-like terms.

The main objective of this work is to derive, from a microscopic starting point, an effective equation of motion for the nanomagnet magnetization that accounts for surface anisotropy and the resulting spin misalignment. 
The approach is based on the spin coherent-state path integral for a many-spin nanomagnet, combined with a controlled expansion in the transverse spin-misalignment field and an adiabatic elimination of fast modes.

The paper is organized as follows.
In Sec.~\ref{sec:PIES}, the path integral formulation in Euclidean space is summarized, and the spin coherent-state representation is introduced.
In Sec.~\ref{sec:Hamiltonian}, the atomistic Hamiltonian of a nanomagnet—including core and surface anisotropy—is presented, and the corresponding continuum action for the magnetization density is derived.
Sec.~\ref{sec:ExtendedLLE} is dedicated to the decomposition of the local magnetization into a slowly varying macrospin and transverse fluctuations. The action is then expanded to quadratic order in the latter, and a closed equation of motion for the macrospin is derived.
In Sec.~\ref{sec:BoundaryEffects}, the analysis focuses on the dominant surface contribution for a spherical nanoparticle, with the leading correction to the effective field explicitly evaluated. The resulting structure of the extended Landau--Lifshitz equation is subsequently discussed.
In Sec.~\ref{sec:Discussion}, the practical relevance of these results—particularly for ferromagnetic resonance measurements—is examined. The underlying adiabatic-elimination mechanism is argued to generalize to other systems with slow/fast variable separation. A comparison is also made with existing inertial and effective one-spin-particle treatments of nutation.
For completeness, explicit relations between the spin Berry phase and curvature are provided in Appendix~\ref{app:Spin_Berry_phase_and_curvature}. The derivation of the standard Landau--Lifshitz equation from the continuum action is reviewed in Appendix~\ref{app:LLE}, and the details of the evaluation of the surface-induced term involving $\pmb{\psi}\times\partial_{t}\pmb{\psi}$ are given in Appendix~\ref{app:ThirdTerm}.

\section{Spin path integral and effective action}
\label{sec:PIES}

The path integral formulation for quantum spin systems is briefly summarized, with emphasis on the Euclidean (imaginary-time) representation and the emergence of the Berry-phase term. This provides the foundation for constructing the effective action for a many-spin nanomagnet.

\subsection{Imaginary-time formalism and path integral}

The path integral method unifies quantum mechanics, quantum field theory, and statistical mechanics, making it a foundational tool in modern theoretical physics~\cite{dasFieldTheoryPath2019}. 
Let $\mathcal{H}$ denote the (real-valued) Hamiltonian of a quantum system in $D$ spatial dimensions.
In the path integral formalism, the time-evolution operator $\exp(-i\mathcal{H}t/\hbar)$ produces an oscillating integrand, so that the path integral must be understood in the sense of distributions \cite{wipfStatisticalApproachQuantum2013}.
Suppressing these oscillations yields a well-defined path integral in imaginary time.
Thus a Wiener measure can be constructed, allowing integration over all paths.
The transformation from real to imaginary time is achieved by a Wick rotation (replacing time $t$ with $-i\tau$), so that one can work with the Euclidean path integral throughout and switch back to real time only at the end ($\tau \to it$).
As a consequence, Wiener's path integral corresponds to Feynman's path integral for imaginary time and fully describes quantum systems in thermal equilibrium with a heat bath at fixed temperature.
The identifications $\frac{i t}{\hbar}=\frac{\tau}{\hbar}=\frac{1}{k_\mathrm{B}T}\equiv\beta$ then produce the desired partition function
\begin{align} 
  Z&=\Tr e^{-\beta\mathcal{H}}, 
\end{align} 
which can be written as a path integral over trajectories in configuration space, with periodic imaginary-time boundary conditions.

The standard construction proceeds by discretizing the interval $\left[0,\beta\right]$ into $N$ steps of size $\delta\tau/\hbar$ and using the Trotter formula,
\begin{align}
  Z&=\Tr e^{-\beta\mathcal{H}}=\lim_{\substack{N\to\infty\\\delta\tau\to0\\N\delta\tau=\beta\hbar}}
  \left(\Tr e^{-\frac{\delta\tau}{\hbar}\mathcal{H}}\right)^{N}.
  \label{eq:Trotter1}
\end{align}
By inserting the identity resolution between the factors $e^{-\frac{\delta\tau}{\hbar}\mathcal{H}}$ and taking the appropriate limits, one arrives at the path integral representation
\begin{align}
  Z&=\int_{\alpha(0)=\alpha(\beta)}\mathcal{D}\alpha(\tau)\,\exp \left(-S_\mathbf{E}[\alpha(\tau)]\right),
  \label{eq:PI-PartFunct}
\end{align}
where $S_\mathbf{E}$ is the Euclidean action functional for all configurations $\alpha$ on which $\mathcal{H}$ depends, as shown in Ref.\cite{wipfStatisticalApproachQuantum2013}.

For spin systems, a convenient choice of basis states $\ket{\alpha}$ is provided by spin coherent states \cite{perelomovGeneralizedCoherentStates1977}. 
These are denoted by $\ket{\mathbf{m}}$, where $\mathbf{m}$ is a unit vector on the two-sphere $S^{2}$. 
For $\mathbf{m}=\mathbf{e}_{z}$ one has $\ket{\mathbf{m}=\mathbf{e}_{z}}=\ket{\uparrow}$, and general coherent states are obtained by applying a suitable spin rotation. 
In spherical coordinates $\mathbf{m}=(\sin\theta\cos\varphi,\sin\theta\sin\varphi,\cos\theta)$.

In terms of spin coherent states, the partition function can be written as \cite{altlandCondensedMatterField2023}
\begin{align}
  Z&=\int\mathcal{D}\mathbf{m}(\tau)\,e^{-\mathcal{S}_{\mathrm{E}}[\mathbf{m}]},
  \label{eq:ZvsSE}
\end{align}
with the Euclidean action
\begin{align}
  \mathcal{S}_{\mathrm{E}}&=\int_{0}^{\beta\hbar}d\tau\,\braket{\mathbf{m}(\tau)|\frac{d}{d\tau}\mathbf{m}(\tau)}
  +\frac{1}{\hbar}\int_{0}^{\beta\hbar}d\tau\,\bra{\mathbf{m}(\tau)}\mathcal{H}\ket{\mathbf{m}(\tau)}.
  \label{eq:EuclideanAction}
\end{align}

The first term in the Euclidean action represents the Berry phase, a geometric phase factor that arises from the system's evolution along a closed path in parameter space~\cite{mooreCalculationNonadiabaticBerry1991}. 
This term is purely imaginary and modifies the path integral's weight.
The expression for $\mathcal{S}_{\mathrm{E}}$ can be simplified by symmetrizing it: since the first term is purely imaginary, its complex conjugate is added and the sum is divided by two.
This reads:
\begin{align}
  \begin{aligned}
  \mathcal{S}_{\mathrm{E}}&=\int_{0}^{\beta\hbar}d\tau\,\left(\frac{1}{2}\left(\braket{\mathbf{m}(\tau)|\frac{d}{d\tau}\mathbf{m}(\tau)}-\braket{\frac{d}{d\tau}\mathbf{m}(\tau)|\mathbf{m}(\tau)}\right)
  +\frac{1}{\hbar}\bra{\mathbf{m}(\tau)}\mathcal{H}\ket{\mathbf{m}(\tau)}\right)\\
  &\equiv\int_{0}^{\beta\hbar}d\tau\,L(\tau)
  \end{aligned}
  \label{eq:EuclideanActionSymmetrized}
\end{align}
where $L(\tau)$ is the corresponding Lagrangian~\cite{sakaiPrinciplesQuantumMechanics2005}.

An alternative expression for the Lagrangian may be obtained from Eq.\eqref{eq:EuclideanAction} as:
\begin{align}
  L(\tau)\equiv\bra{\mathbf{m}(\tau)}\frac{d}{d\tau}+\frac{\mathcal{H}}{\hbar}\ket{\mathbf{m}(\tau)}
\end{align}

This form is obtained by adding the total time-derivative $\frac{d}{d\tau}\braket{\mathbf{m}(\tau)|\mathbf{m}(\tau)}$ to Eq. \eqref{eq:EuclideanActionSymmetrized}, which does not alter the equations of motion.
In particular, this substitution renders the Lagrangian complex-valued, while yielding results consistent with those derived from Eq.~\eqref{eq:EuclideanActionSymmetrized}.
Since the expression is restricted to first-order time derivatives of the state vector, the resulting Euler--Lagrange equations of motion reduce to:
\begin{align}
  \hfill\qquad\frac{d}{d\tau}\left(\frac{\partial L}{\partial\ket{\frac{d}{d\tau}\mathbf{m}(\tau)}}\right)-\frac{\partial L}{\partial\ket{\mathbf{m}(\tau)}}=0,&\qquad\frac{d}{d\tau}\left(\frac{\partial L}{\partial\bra{\frac{d}{d\tau}\mathbf{m}(\tau)}}\right)-\frac{\partial L}{\partial\bra{\mathbf{m}(\tau)}}=0
  \label{eq:EOMs}
\end{align}
which in turn lead to the equation of motion for the ket:
\begin{align}
    -\frac{d}{d\tau}\ket{\mathbf{m}(\tau)}&=\frac{\mathcal{H}}{\hbar}\ket{\mathbf{m}(\tau)}\;\iff\;
    i\hbar\frac{d}{dt}\ket{\mathbf{m}(t)}=\mathcal{H}\ket{\mathbf{m}(t)},
\end{align}
from which the Schrödinger equation follows.

This analysis employs a \textit{continuous-time regularization} for the phase space path integral.
This method ensures proper convergence and addresses potential singularities in the spin coherent-state representation while preserving the physical content of the Berry phase term, which is essential for capturing the correct dynamics of the spin system~\cite{arafuneGardenQuantaEssays2003,wilsonBreakdownCoherentState2011}.

\subsection{Spin coherent states and Wess--Zumino--Witten term}
\label{subsec:scs_wzw}

The first term in Eq.~\eqref{eq:EuclideanAction} represents the Berry phase, or Wess--Zumino--Witten (WZW) action $-iS\mathcal{S}_{\mathrm{WZW}}[\mathbf{m}]$~\citep{fradkinFieldTheoriesCondensed2015}, which by Stokes' theorem transforms into a surface integral:
\begin{align}
  \int_0^{\beta\hbar}\,\braket{\mathbf{m}(\tau)|\frac{d\mathbf{m}(\tau)}{d\tau}}d\tau &=\int_{\bm{m}(0)}^{\bm{m}(\beta\hbar)}\,\braket{\mathbf{m}|d\mathbf{m}}=\int_0^{\beta\hbar}\int_{0}^1\braket{d\mathbf{m}(\tau,u)|\wedge|d\mathbf{m}(\tau,u)},
\end{align}
where $\mathbf{m}(\tau,u)$ is a smooth extension of the physical trajectory $\mathbf{m}(\tau)$ into an auxiliary dimension $u\in[0,1]$, satisfying periodic boundary conditions in $\tau$.
The field $\mathbf{m}(\tau,u)$ interpolates between a reference configuration at $u=0$ and the classical path at $u=1$, such that the integral of the symplectic 2-form $\braket{d\mathbf{m}|\wedge|d\mathbf{m}}$ corresponds to the geometric area enclosed by the trajectory~\cite{avronHomotopyQuantizationCondensed1983}.

Since the manifold on which $\mathbf{m}$ evolves is compact, this term reads [see Appendix~\ref{app:Spin_Berry_phase_and_curvature}]
\begin{align}
  \mathcal{S}_{\mathrm{WZW}}[\mathbf{m}]&=\int_{0}^{\beta\hbar}d\tau\int_{0}^{1}du
  \mathbf{m}(\tau,u)\cdot
  \left[\frac{\partial\mathbf{m}}{\partial\tau}\times\frac{\partial\mathbf{m}}{\partial u}\right],
  \label{eq:S_WZW_spin}
\end{align}

The full Euclidean action for a generic spin Hamiltonian $\mathcal{H}$ is then
\begin{align}
  \mathcal{S}_{\mathrm{E}}&=-iS\mathcal{S}_{\mathrm{WZW}}[\mathbf{m}]
  +\frac{1}{\hbar}\int_{0}^{\beta\hbar}d\tau\,\bra{\mathbf{m}(\tau)}\mathcal{H}\ket{\mathbf{m}(\tau)}.
  \label{eq:EuclideanAction_v2}
\end{align}

In the following, $\mathcal{H}$ is specified for a nanomagnet comprising exchange, Zeeman, and core/surface anisotropy contributions. Eq.~\eqref{eq:EuclideanAction_v2} represents the Euclidean action useful in the thermal coherent-state path integral. Conversely, the semiclassical magnetization dynamics are derived through analytic continuation to real time, with the resulting magnetic action denoted as $\mathcal{S}_{\mathrm{M}}$. Under this convention, the WZW contribution to $\mathcal{S}_{\mathrm{M}}$ is rendered real, while the factor $-i$ remains in the Euclidean action $\mathcal{S}_{\mathrm{E}}$. This distinction accounts for the absence of an explicit factor $i$ in the WZW terms presented in Eqs.~\eqref{eq:WZW_MSP_Continuum}, \eqref{eq:Action_n0}, and \eqref{eq:Action_S2}.

\section{Many-spin nanomagnet: Hamiltonian and continuum limit}
\label{sec:Hamiltonian}

\subsection{Atomistic Hamiltonian with core and surface anisotropy}

In the atomistic, many-spin approach, a nanomagnet is described by the Hamiltonian
\begin{equation}
\mathcal{H}=\mathcal{H}_{E}+\mathcal{H}_{Z}+\mathcal{H}_{c}+\mathcal{H}_{s},
\label{Ham}
\end{equation}
where
\begin{equation}
\mathcal{H}_{E}=-\frac{1}{2}\sum_{ij}J_{ij}\,\mathbf{s}_{i}\cdot\mathbf{s}_{j}
\label{HamE}
\end{equation}
is the exchange energy,
\begin{equation}
\mathcal{H}_{Z}=-\mu_{a}\mathbf{H}\cdot\sum_{i=1}^{\mathcal{N}}\mathbf{s}_{i}
\label{eq:HamZ}
\end{equation}
is the Zeeman energy\footnote{Here $H$ denotes the field expressed in Tesla, i.e., $\mu_0$ times the field strength in A/m.}, with $\mu_{a}=n_{a}\mu_{B}$ the atomic magnetic moment, and
\begin{equation}
\mathcal{H}_{c}=-K_{c}\sum_{i\in\mathrm{core}}s_{iz}^{2}
\label{eq:HamCore}
\end{equation}
is the uniaxial core anisotropy with easy axis along the $z$ direction.
The surface anisotropy is described by the N\'{e}el model~\citep{neelAnisotropieMagnetiqueSuperficielle1954,garaninSurfaceContributionAnisotropy2003,kachkachiSurfaceinducedCubicAnisotropy2006},
\begin{equation}
\mathcal{H}_{s}=\frac{K_{s}}{2}\sum_{i\in\mathrm{surface}}\sum_{j=1}^{z_{i}}
\bigl(\mathbf{s}_{i}\cdot\mathbf{e}_{ij}\bigr)^{2},
\label{eq:NSA}
\end{equation}
where the sum over $j$ runs over the $z_{i}$ nearest neighbours
of site $i$ and $\mathbf{e}_{ij}$ is the unit vector connecting
$i$ and $j$. The factor $1/2$ avoids double counting of the bonds.

The total Hamiltonian can be written as
\begin{equation}
\mathcal{H}=-\frac{1}{2}\sum_{ij}J_{ij}\,\mathbf{s}_{i}\cdot\mathbf{s}_{j}
-\mu_{a}\mathbf{H}\cdot\sum_{i=1}^{\mathcal{N}}\mathbf{s}_{i}
-K_{c}\sum_{i\in\mathrm{core}}s_{iz}^{2}
+\frac{K_{s}}{2}\sum_{i,j}^{\mathcal{N}}(\mathbf{s}_{i}\cdot\mathbf{e}_{ij})^{2},
\label{eq:TotalHam}
\end{equation}
from which the local effective field at site $i$ follows as
\begin{equation}
\mathbf{H}_{i,\mathrm{eff}}\equiv-\frac{1}{\mu_{a}}\frac{\delta\mathcal{H}}{\delta\mathbf{s}_{i}}
=\mathbf{H}+\frac{2K_{c}}{\mu_{a}}s_{iz}\,\mathbf{e}_{z}
-\frac{K_{s}}{\mu_{a}}\sum_{j=1}^{z_{i}}(\mathbf{s}_{i}\cdot\mathbf{e}_{ij})\mathbf{e}_{ij}
+\frac{1}{\mu_{a}}\sum_{j}J_{ij}\,\mathbf{s}_{j}.
\label{eq:TotalEffField}
\end{equation}

\subsection{Continuum magnetization field and anisotropy functional}

For the purposes of the present derivation, a transition to a continuum description is performed, wherein the magnetic configuration is represented by a magnetization density $\mathbf{M}(\mathbf{r},t)$, a field of constant magnitude $M_{0}$. A unit vector field is subsequently defined as
\begin{align}
  \label{eq:mr}
  \mathbf{m}(\mathbf{r},t)&=\mathbf{M}(\mathbf{r},t)/M_{0}.
\end{align}
The correspondence between the discrete and continuum representations is established by the relation
\begin{align}
  \label{eq:Mr}
  \mathbf{M}(\mathbf{r},t)&=\sum_{i}\frac{\bm{\mu}_{i}(t)}{v_{0}},
\end{align}
and the summation is carried out over all sites within a small volume $v_{0}$ surrounding $\mathbf{r}$, within which the moments are assumed to be uniform.

In the continuum limit the Hamiltonian \eqref{eq:TotalHam} becomes
\begin{align}
  \mathcal{H}&=-\frac{S^{2}}{2}\sum_{\langle\mathbf{r},\mathbf{r}'\rangle} J(\mathbf{r},\mathbf{r}')\,
  \mathbf{m}(\mathbf{r},t)\cdot\mathbf{m}(\mathbf{r}',t)
  -\mu_{a}S\sum_{\mathbf{r}}\mathbf{H}\cdot\mathbf{m}(\mathbf{r},t) 
  -S^{2}\sum_{\mathbf{r}}\mathcal{A}[\mathbf{m}(\mathbf{r},t)],
\end{align}
where $\mathcal{A}$ is the on-site anisotropy functional. 
It can be written as
\begin{align}
  \mathcal{A}\bigl[\mathbf{m}(\mathbf{r},t)\bigr]&=
  \begin{cases}
    K_{c}[\mathbf{m}(\mathbf{r},t)\cdot\mathbf{e}_{z}]^{2}, & \mathbf{r}\in\mathrm{core},\\[3pt]
    -\dfrac{1}{2}K_{s}\displaystyle\sum_{\boldsymbol{\delta}}
    [\mathbf{m}(\mathbf{r},t)\cdot\mathbf{e}(\mathbf{r},\boldsymbol{\delta})]^{2},
    & \mathbf{r}\in\mathrm{surface},
  \end{cases}
  \label{eq:AnisFunctional_v0}
\end{align}
or equivalently as a quadratic form
\begin{align}
  \mathcal{A}[\mathbf{m}(\mathbf{r},t)]
  &=\sum_{\alpha\beta}g_{\alpha\beta}(\mathbf{r})\,m^{\alpha}(\mathbf{r},t)m^{\beta}(\mathbf{r},t),
  \label{eq:AnisFunctional_v1}
\end{align}
with anisotropy tensor
\begin{align}
  g_{\alpha\beta}(\mathbf{r})&=
  \begin{cases}
    K_{c}e_{z}^{\alpha}e_{z}^{\beta}, & \mathbf{r}\in\mathrm{core},\\[3pt]
    -\dfrac{1}{2}K_{s}\displaystyle\sum_{\boldsymbol{\delta}}
    e_{\alpha}(\mathbf{r},\boldsymbol{\delta})e_{\beta}(\mathbf{r},\boldsymbol{\delta}),
    & \mathbf{r}\in\mathrm{surface},
  \end{cases}
  \label{eq:AnisotropyTensor}
\end{align}
where $\mathbf{e}(\mathbf{r},\boldsymbol{\delta})$ is the unit vector from site $\mathbf{r}$ to its neighbour at distance $\delta$.

Using the standard replacement $\sum_{\mathbf{r}}\to (1/a_{0}^{3})\int d\mathbf{r}$ and the constraint $\mathbf{m}^{2}=1$, the exchange term can be written, up to a constant, as
\begin{align*}
  \int_{0}^{T}\!dt\sum_{\langle\mathbf{r},\mathbf{r}'\rangle}J(\mathbf{r},\mathbf{r}') \,\mathbf{m}(\mathbf{r},t)\cdot\mathbf{m}(\mathbf{r}',t) 
  &\longrightarrow \frac{1}{a_{0}}\int_{0}^{T}\!dt\int d\mathbf{r}\, J(\mathbf{r},\mathbf{r}')\bigl[\nabla\mathbf{m}(\mathbf{r},t)\bigr]^{2}.
\end{align*}
Collecting contributions after analytic continuation to real time, the continuum magnetic action reads
\begin{align}
  \begin{aligned}
    \mathcal{S}_{\mathrm{M}}[\mathbf{m}]
    =&\frac{S}{a_{0}^{3}}\int d\mathbf{r}\,\mathcal{S}_{\mathrm{WZW}}[\mathbf{m}(\mathbf{r},t)]\\
    &+\frac{S^{2}}{2a_{0}}\int_{0}^{T}\!dt\int d\mathbf{r}\,
    J(\mathbf{r},\mathbf{r}')\bigl[\nabla\mathbf{m}(\mathbf{r},t)\bigr]^{2}\\
    &+\frac{\mu_{a}S}{a_{0}^{3}}\int_{0}^{T}\!dt\int d\mathbf{r}\,\mathbf{H}\cdot\mathbf{m}(\mathbf{r},t)
    +\frac{S^{2}}{a_{0}^{3}}\int_{0}^{T}\!dt\int d\mathbf{r}\,\mathcal{A}[\mathbf{m}(\mathbf{r},t)].
  \end{aligned}
  \label{eq:WZW_MSP_Continuum}
\end{align}

As discussed at the end of Section \ref{subsec:scs_wzw}, here and in the following, $\mathcal{S}_{\mathrm{M}}$ denotes the real-time magnetic action obtained from the Euclidean coherent-state action $\mathcal{S}_{\mathrm{E}}$ after analytic continuation; it should therefore not be confused with the Euclidean weight appearing in Eq.~\eqref{eq:ZvsSE}. From this action, the undamped Landau--Lifshitz equation can be recovered, as is done in Appendix~\ref{app:LLE}.

\section{Macrospin--fluctuation decomposition and perturbation theory}
\label{sec:ExtendedLLE}

\subsection{Decomposition into macrospin and transverse fluctuations}

\begin{figure}[htbp]
  \begin{centering}
  \includegraphics[angle=-45,width=5.5cm]{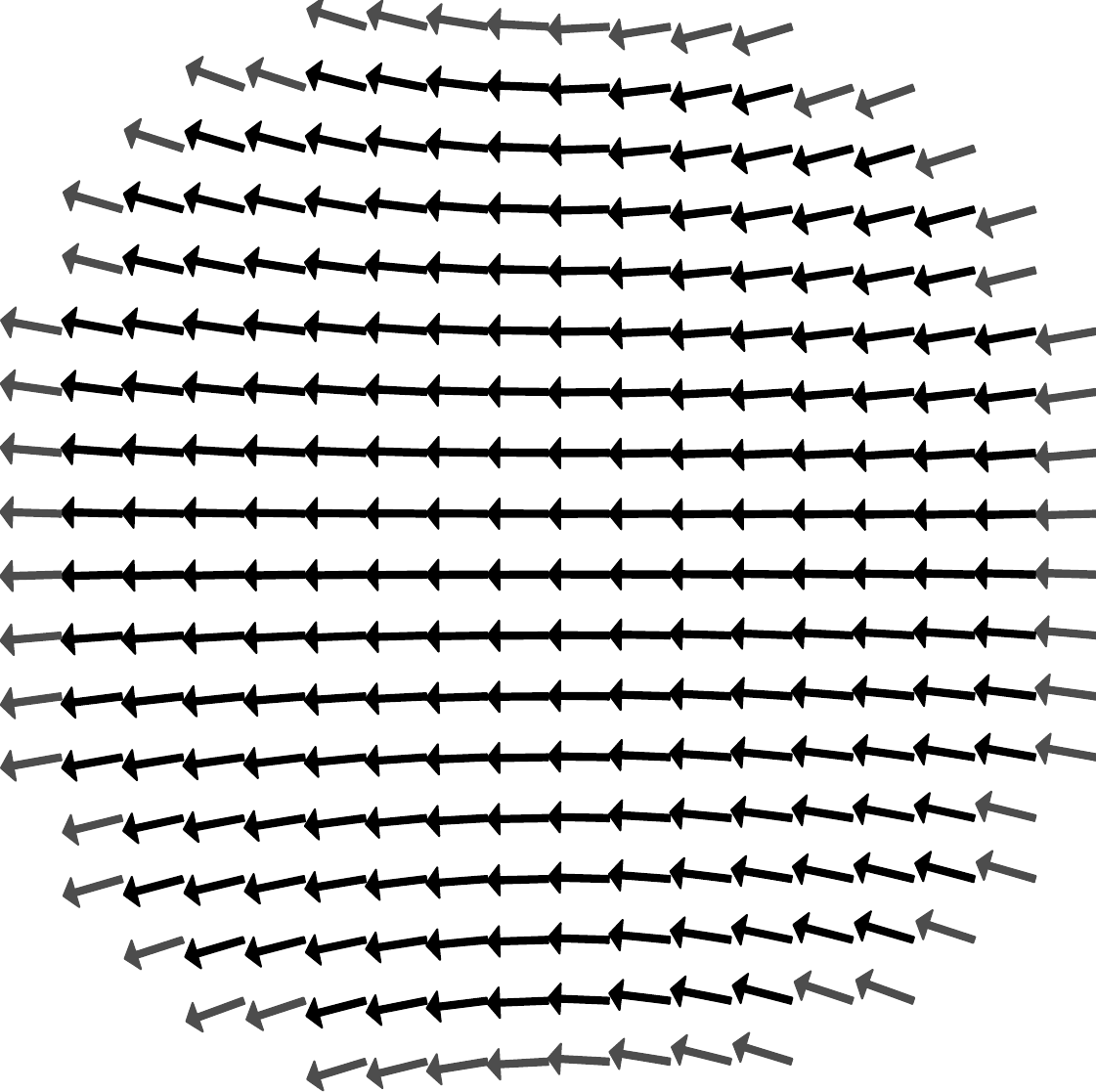}
  \par
  \end{centering}
  \caption{Magnetic structure of a spherical nanoparticle of linear size $N=20$, showing atoms in the plane $z=0$. The spin configuration deviates from the homogeneous state near the surface.}
  \label{fig:nsa-structure}
\end{figure}

In a nanomagnet with strong surface anisotropy the magnetic state cannot be treated as homogeneous: boundary effects induce spin misalignment, and an atomistic description is needed to capture the resulting nonuniformity.
As a consequence, the greatest deviations from a collinear state occur near the nanomagnet's boundary (see, for example, Fig.~\ref{fig:nsa-structure}).

To account for these effects, the instantaneous and local magnetization density $\mathbf{m}(\mathbf{r},t)$ is decomposed into a homogeneous macrospin component and a transverse fluctuation field~\citep{garaninMagnetizationReversalInternal2009},
\begin{align}
  \mathbf{m}(\mathbf{r},t)&=\mathbf{m}_{0}(t)\sqrt{1-\|\bm{\psi}(\mathbf{r},t)\|^{2}}+\bm{\psi}(\mathbf{r},t),
  \label{eq:ExactLinarization-1}
\end{align}
where $\mathbf{m}_{0}\cdot\bm{\psi}=0$, $\|\mathbf{m}(\mathbf{r},t)\|=1$, and the condition
\begin{align}
  \int_{V}d^{3}r\,\bm{\psi}(\mathbf{r},t)&=0, \quad \forall t\in\mathbb{R}
  \label{eq:Intpsizero}
\end{align}
is imposed. Here, $\mathbf{m}_{0}$ denotes the orientation of the net magnetization, defined as a unit vector such that $\|\mathbf{m}_0\|=1$.

For small spin misalignment, the magnetization density deviates only slightly from $\mathbf{m}_{0}$, allowing a perturbative treatment in $\boldsymbol{\psi}$. Under the assumption $|\psi_{\alpha}|\ll 1$, Eq.~\eqref{eq:ExactLinarization-1} is expanded to second order in $\boldsymbol{\psi}$, yielding
\begin{align}
  \mathbf{m}(\mathbf{r},t)&\simeq\mathbf{m}_{0}(t)+\delta\mathbf{m}(\mathbf{r},t),
  \label{eq:SpinVector-expand}
\end{align}
where the deviation is given by
\begin{align}
  \delta\mathbf{m}(\mathbf{r},t)&\equiv\bm{\psi}(\mathbf{r},t)
  -\frac{1}{2}\bm{\psi}^{2}(\mathbf{r},t)\,\mathbf{m}_{0}(t)
  -\frac{1}{8}\bm{\psi}^{4}(\mathbf{r},t)\,\mathbf{m}_{0}(t).
  \label{eq:SpinDeficit}
\end{align}

$\mathbf{m}_{0}$ is the net magnetic moment of the nanomagnet (macrospin) whose dynamics is coupled to that of the transverse fluctuation field $\bm{\psi}$.

\subsection{Expansion of the action and quadratic form in \texorpdfstring{$\bm{\psi}$}{psi}}

Inserting the decomposition \eqref{eq:SpinVector-expand} into the action~\eqref{eq:WZW_MSP_Continuum} and expanding up to quadratic order in $\bm{\psi}$ yields
\begin{align}
  \mathcal{S}_{\mathrm{M}}[\mathbf{m}] &=\mathcal{S}_{\mathrm{M}}^{(0)}[\mathbf{m}_{0}] +\mathcal{S}_{\mathrm{M}}^{(2)}[\bm{\psi},\mathbf{m}_{0}],
  \label{eq:Action_S0S2}
\end{align}
where
\begin{align}
  \begin{aligned}
  \mathcal{S}_{\mathrm{M}}^{(0)}[\mathbf{m}_{0}]
  =&\frac{\mu_{a}S}{a_{0}^{3}}\int_{0}^{T}\!dt\int d\mathbf{r}\,\mathbf{H}\cdot\mathbf{m}_{0}(t)
  +\frac{S^{2}}{a_{0}^{3}}\int_{0}^{T}\!dt\int d\mathbf{r}\,\mathcal{A}[\mathbf{m}_{0}]\\
  &\quad+\frac{S}{a_{0}^{3}}\int d\mathbf{r}
  \int_{0}^{T}\!dt\int_{0}^{1}\!du\,
  \mathbf{m}_{0}(t,u)\cdot\bigl(\partial_{u}\mathbf{m}_{0}\times\partial_{t}\mathbf{m}_{0}\bigr)
  \end{aligned}
  \label{eq:Action_n0}
\end{align}
is the unperturbed macrospin action, and
\begin{align}
  \begin{aligned}
    \mathcal{S}_{\mathrm{M}}^{(2)}[\bm{\psi},\mathbf{m}_{0}]
    &=\frac{JS^{2}}{2a_{0}}\int d\mathbf{r}\int_{0}^{T}\!dt\,
    \bigl[\nabla\bm{\psi}(\mathbf{r},t)\bigr]^{2}
    -\frac{\mu_{a}S}{2a_{0}^{3}}\int_{0}^{T}\!dt\int d\mathbf{r}\,
    (\mathbf{H}\cdot\mathbf{m}_{0})\,\bm{\psi}^{2}
    \\
    &+\frac{S^{2}}{a_{0}^{3}}\int d\mathbf{r}\int_{0}^{T}\!dt\,
    \sum_{\alpha\beta}g_{\alpha\beta}
    \bigl(\psi^{\alpha}\psi^{\beta}-\bm{\psi}^{2}m_{0}^{\alpha}m_{0}^{\beta}\bigr)
    \\
    &-\frac{S}{a_{0}^{3}}\int d\mathbf{r}\int_{0}^{T}\!dt\,
    \mathbf{m}_{0}(t)\cdot\bigl(\bm{\psi}\times\partial_{t}\bm{\psi}\bigr)
  \end{aligned}
  \label{eq:Action_S2}
\end{align}
is the quadratic correction. 
In the derivation of these expressions, the following relations were used:

\begin{itemize}
  \item{Terms linear in $\bm{\psi}$ vanish upon spatial integration because of the constraint \eqref{eq:Intpsizero}.}
  \item{Terms of the form $\bm{\psi}^{2}\,\mathbf{m}_{0}\cdot(\partial_{u}\mathbf{m}_{0}\times\partial_{t}\mathbf{m}_{0})$ are higher order within the combined small-misalignment and slow-macrospin expansion, and are therefore neglected.}
  \item{The WZW contribution is expanded to quadratic order in the transverse field. The leading fluctuation-dependent Berry term retained in the quadratic action is the physical-boundary contribution
  \[
    -\frac{S}{a_{0}^{3}}\int d\mathbf{r}\int_{0}^{T}\!dt\,
    \mathbf{m}_{0}(t)\cdot\bigl(\bm{\psi}\times\partial_{t}\bm{\psi}\bigr).
  \]
  Other terms involving the auxiliary WZW extension, such as
  \[
    \mathbf{m}_{0}\cdot
    \bigl(\partial_{u}\bm{\psi}\times\partial_{t}\bm{\psi}\bigr),
  \]
  are not assumed to vanish identically. Rather, they are discarded here as subleading extension-dependent contributions within the present adiabatic truncation. Keeping them consistently would require specifying the full $u$-dependent extension of the transverse field and retaining the corresponding higher-order corrections to the transverse-mode dynamics.}
  \item{At $u=0$ one has $\mathbf{m}_{0}(t,0)=\mathbf{m}_{0}$, implying $\bm{\psi}=\mathbf{0}$, while at $u=1$ one recovers the physical configuration $\mathbf{m}_{0}(t)$.}
  \item{The spatial variation of exchange $J(\bm{r},\bm{r}')$ is considered uniform.}
\end{itemize}

\subsection{Effective equation of motion for the macrospin}

To obtain the equation of motion for $\mathbf{m}_{0}$, the constraint term enforcing $\mathbf{m}_{0}^{2}=1$ is added,
\begin{equation}
\mathcal{S}_{\mathrm{L}}[\lambda,\mathbf{m}_{0}]
=\frac{1}{2}\int_{0}^{T}\!dt\int d\mathbf{r}\,
\lambda(\mathbf{r},t)\left[\mathbf{m}_{0}^{2}(\mathbf{r},t)-1\right],
\label{eq:ConstraintAction}
\end{equation}
and the total action
\[
\mathcal{S}_{\mathrm{T}}[\mathbf{m}_{0},\boldsymbol{\psi}]
=\mathcal{S}_{\mathrm{M}}[\mathbf{m}_{0},\boldsymbol{\psi}]
+\mathcal{S}_{\mathrm{L}}[\lambda,\mathbf{m}_{0}]
\]
is required to be stationary under variations of $\mathbf{m}_{0}$. The relevant functional derivatives are detailed in the text; only the result is summarized here.

The functional derivative of the anisotropy functional with respect to $\mathbf{m}_{0}$ reads
\begin{align}
  \frac{\delta\mathcal{A}[\mathbf{m}_{0}]}{\delta m_{0}^{\gamma}} &
  =2\sum_{\beta}g_{\gamma\beta}(\mathbf{r})\,m_{0}^{\beta}. 
  \label{eq:RAM-Continuous_m0}
\end{align}
After algebraic rearrangement, the functional derivative of the action is found to be
\begin{align*}
  \frac{\delta\mathcal{S}_{\mathrm{M}}}{\delta m_{0}^{\alpha}}
  &=S(\mathbf{m}_{0}\times\partial_{t}\mathbf{m}_{0})_{\alpha}
  +\mu_{a}SH_{\alpha}
  +2S^{2}\int d\mathbf{r}'\sum_{\beta}g_{\alpha\beta}(\mathbf{r})m_{0}^{\beta}
  -\frac{\mu_{a}S}{2}H_{\alpha}\,\overline{\bm{\psi}^{2}}\\
  &\quad-2S^{2}\int d\mathbf{r}'\left[\sum_{\beta}g_{\alpha\beta}(\mathbf{r})m_{0}^{\beta}\right]\bm{\psi}^{2}(\mathbf{r})
  -S\int d\mathbf{r}'\bigl(\bm{\psi}\times\partial_{t}\bm{\psi}\bigr)_{\alpha}
  +\lambda m_{0}^{\alpha},
\end{align*}
where the spatial average is defined as
\begin{align*}
  \overline{\bm{\psi}^{2}}&=
   \frac{1}{a_{0}^{3}}\int d\mathbf{r}\,\bm{\psi}^{2}(\mathbf{r}) \equiv\int d\mathbf{r}'.
\end{align*}

Setting $\delta\mathcal{A}[\mathbf{m}_{0}]/\delta m_{0}^{\gamma}$ to zero and solving for the Lagrange multiplier $\lambda$ by taking the scalar product with $\mathbf{m}_{0}$, one eventually obtains the equation of motion in the form
\begin{align}
  \mathbf{m}_{0}\times\partial_{t}\mathbf{m}_{0}
  +\mathbf{m}_{0}\times\bigl(\mathbf{H}_{\mathrm{eff}}\times\mathbf{m}_{0}\bigr)&=0,
\end{align}
or equivalently
\begin{align}
  \begin{aligned}
    \partial_{t}\mathbf{m}_{0}
    &=\mathbf{m}_{0}\times\mathbf{H}_{\mathrm{eff}}\\
    &=\mathbf{m}_{0}\times\left[
    \mu_{a}\left(1-\frac{1}{2}\overline{\bm{\psi}^{2}}\right)\mathbf{H}
    +\int d\mathbf{r}'\,(1-\bm{\psi}^{2})\mathbf{H}_{A}
    -\int d\mathbf{r}'\left(\bm{\psi}\times\partial_{t}\bm{\psi}\right)\right],
  \end{aligned}
  \label{eq:EM_m0_RHSinPsi}
\end{align}
where the anisotropy field is
\begin{align*}
  \mathbf{H}_{A,\gamma}&=2S\sum_{\beta}g_{\gamma\beta}(\mathbf{r})\,m_{0}^{\beta}.
\end{align*}

Equation~\eqref{eq:EM_m0_RHSinPsi} becomes a closed equation for $\mathbf{m}_{0}$ once $\bm{\psi}$ is expressed in terms of $\mathbf{m}_{0}$ and $\partial_{t}\mathbf{m}_{0}$, as discussed next.

\section{Structure of the extended LLE: adiabatic approximation}
\label{sec:BoundaryEffects}

\subsection{Adiabatic solution and the effective-field correction}

As discussed above, the transverse fluctuation field $\bm{\psi}(\mathbf{r},t)$ is induced by surface anisotropy and other inhomogeneities. 
In previous works \cite{garaninSurfaceContributionAnisotropy2003, kachkachiEffectsSpinNoncollinearities2007, bastardisMagnetizationNutationInduced2018, adamsLowfrequencySignatureMagnetization2024}, it was found that the characteristic time scale of the dynamics of $\bm{\psi}$ is much shorter than that of the macrospin $\mathbf{m}_{0}$, so that an adiabatic approximation is plausible: the transverse modes are assumed to rapidly adjust to the instantaneous macrospin configuration.

In this adiabatic approach, $\bm{\psi}(\mathbf{r},t)$ is replaced by its static solution expressed in terms of $\mathbf{m}_{0}$, which is obtained by solving the Helmholtz equation for $\bm{\psi}$ using a Green's-function method~\citep{garaninSurfaceContributionAnisotropy2003,kachkachiEffectsSpinNoncollinearities2007,adamsSpatialMagnetizationProfile2023}.
The equation for $\bm{\psi}$ can be recovered from the quadratic action~\eqref{eq:Action_S2} by applying the Euler--Lagrange equation $\delta\mathcal{L}/\delta\psi_{\alpha}-\partial_{\mu}\delta\mathcal{L}/\delta(\partial_{\mu}\psi_{\alpha})=0$. 
The structure of the solution has been detailed in Ref.~\citep{adamsSpatialMagnetizationProfile2023}.

For a spherical nanomagnet of radius $R$ one introduces the reduced coordinate $\boldsymbol{\xi}=\mathbf{r}/R$. 
The static solution inside the volume and on its boundary can be written as\cite{duffyGreensFunctionsApplications2014, garaninSurfaceContributionAnisotropy2003, adamsSpatialMagnetizationProfile2023}
\begin{align}
  \psi_{\beta}(\boldsymbol{\xi})&=\frac{1}{4\pi}\oint_{\partial V}d^{2}n'\,
  \Sigma_{\beta}(\mathbf{m}_{0},\mathbf{n}')\,\mathcal{G}_{\beta}(\boldsymbol{\xi},\mathbf{n}'),
  \label{eq:GF_Helmholtz_general3}
\end{align}
where $\Sigma_{\beta}(\mathbf{m}_{0},\mathbf{n})=(\nabla_{\xi}\psi_{\beta})\cdot\mathbf{n}$
is the outward normal derivative of $\psi_{\beta}$ at the surface, $\mathbf{n}$ the surface normal, and $\mathcal{G}_{\beta}$ the appropriate Green's function.
Explicit expressions for $\psi_{\beta}$ in terms of position and macrospin orientation, with and without core anisotropy, were obtained in Ref.~\citep{adamsSpatialMagnetizationProfile2023}.

Inserting the adiabatic solution for $\bm{\psi}$ into Eq.~\eqref{eq:EM_m0_RHSinPsi} leads to a closed equation of motion for $\mathbf{m}_{0}$ with an effective field
\begin{align}
  \mathbf{H}_{\mathrm{eff}}&=\mathbf{H}_{\mathrm{eff}}^{(0)}+\delta\mathbf{H}_{\mathrm{eff}},
\end{align}
where
\begin{align*}
  \mathbf{H}_{\mathrm{eff}}^{(0)}&=\mu_{a}\mathbf{H}+\mathbf{H}_{A}
\end{align*}
is the effective field for a homogeneous macrospin, and
\begin{align}
  \delta\mathbf{H}_{\mathrm{eff}}&=
  -\frac{1}{2}\overline{\bm{\psi}^{2}}\,\mu_{a}\mathbf{H} 
  -\int d\mathbf{r}'\,\bm{\psi}^{2}\mathbf{H}_{A} 
  -\int d\mathbf{r}'\left(\bm{\psi}\times\partial_{t}\bm{\psi}\right)
  \label{eq:CorrectionField}
\end{align}
is the correction due to spin misalignment. The first two terms renormalize the Zeeman and anisotropy fields, while the last term contains the time derivative of $\bm{\psi}$ and generates additional dynamical contributions.

The precession frequency, proportional to $\gamma|\mathbf{H}_{\mathrm{eff}}|$, is thus modified compared to that of the homogeneous macrospin, with correction $\delta\mathbf{H}_{\mathrm{eff}}$.
Contributions that are quadratic in $\mathbf{m}_{0}$ and its time derivative arise from the term involving $\partial_{t}\bm{\psi}$, which may be interpreted as an effective damping or nutation contribution, as discussed further below.

This renormalization of the precession frequency is of more than academic interest. Because the magnitude and orientation of the effective field are modified already at order $k_{s}^{2}$, the standard ferromagnetic-resonance (FMR) frequency and lineshape are directly shifted. This signature is measurable with conventional GHz-range FMR spectrometers, as demonstrated in Ref.~\citep{adamsLowfrequencySignatureMagnetization2024}. Unlike the nutation resonance itself—which lies in the sub-THz/THz range and requires dedicated broadband or time-resolved setups—this low-frequency fingerprint of the underlying surface-induced spin misalignment does not require the fast dynamics to be resolved. This point, along with its broader implications, is returned to in Sec.~\ref{sec:Discussion}.

In what follows, the dominant contribution from surface effects is addressed by evaluating the relevant integrals on the boundary of the nanomagnet, while providing the explicit expressions derived in Ref.~\citep{adamsSpatialMagnetizationProfile2023}.

\subsection{Boundary contributions and explicit expressions}

The strongest deviations from a collinear state are observed at the nanomagnet surface, i.e., for $\boldsymbol{\xi}=\mathbf{n}$. Under these conditions, the components of the deviation field $\bm{\psi}$ are approximated by~\citep{adamsSpatialMagnetizationProfile2023}
\begin{align}
  \begin{aligned}
    \psi_{1}(\mathbf{n},\mathbf{m}_{0})
    &\simeq\lambda_{s}\left(1-\frac{\kappa_{1}^{2}}{14}\right)
    \frac{m_{0,x}m_{0,y}}{\sqrt{1-m_{0,z}^{2}}}\,(n_{x}^{2}-n_{y}^{2}),\\
    \psi_{2}(\mathbf{n},\mathbf{m}_{0})
    &\simeq\lambda_{s}\left(1-\frac{\kappa_{2}^{2}}{14}\right)
    \frac{m_{0,z}}{\sqrt{1-m_{0,z}^{2}}}
    \left[(n_{x}^{2}-n_{z}^{2})m_{0,x}^{2}+(n_{y}^{2}-n_{z}^{2})m_{0,y}^{2}\right],
  \end{aligned}
  \label{eq:psi-LargestDev}
\end{align}
where $\mathbf{n}=(\sin\theta\cos\varphi,\sin\theta\sin\varphi,\cos\theta)$
is the surface normal, and
\begin{align*}
  \lambda_{s}=\frac{15}{32}\widetilde{k}_{s}&=\frac{15}{32}\left(\frac{D}{a}\right)k_{s}
\end{align*}
a dimensionless surface anisotropy parameter.

The Helmholtz coefficients $\kappa_1$ and $\kappa_2$ are defined as
\begin{equation}
\left\{ \begin{array}{lll}
\kappa_{1}^{2} & = & \left(\widetilde{\mathbf{h}}_{\mathrm{ext}}\cdot\mathbf{m}_{0}\right)+2\widetilde{k}_{c}\cdot\left(\mathbf{m}_{0}\cdot\mathbf{e}_{A}\right)^{2},\\
\\
\kappa_{2}^{2} & = & \left(\widetilde{\mathbf{h}}_{\mathrm{ext}}\cdot\mathbf{m}_{0}\right)+2\widetilde{k}_{c}\cdot\left[2\left(\mathbf{m}_{0}\cdot\mathbf{e}_{A}\right)^{2}-1\right].
\end{array}\right.\label{eq:HelmholtzCoefficients}
\end{equation}
where
\begin{align}
  \widetilde{k}_{c} & =\frac{1}{2}\left(\frac{D}{a}\right)^{2}k_{c}, & \widetilde{k}_{s} & =\left(\frac{D}{a}\right)k_{s}, & \widetilde{h}_{\mathrm{ext}} & =\frac{1}{2}\left(\frac{D}{a}\right)^{2}h_{\mathrm{ext}},\label{eq:DecoupledHelmholtzEquationCoefficient3}
\end{align}
with $D=2R$ being the diameter of the NM, $k_{c}\equiv K_{c}/J$ and $k_{s}\equiv K_{s}/J$ the (dimensionless) reduced anisotropy constants, and $h_{\mathrm{ext}}\equiv\mu_{a}H_{\mathrm{ext}}/J$ the reduced magnetic field.

Explicit corrections to the effective field are obtained by inserting Eqs.~\eqref{eq:psi-LargestDev} into $\overline{\bm{\psi}^{2}}$, $\int d\mathbf{r}'\,\bm{\psi}^{2}\mathbf{H}_{A}$, and $\int d\mathbf{r}'(\bm{\psi}\times\partial_{t}\bm{\psi})$.
The necessary surface integrals, such as
\begin{align}
  \frac{1}{4\pi}\oint_{\partial V}\left(n_{\alpha}^{2}-n_{\beta}^{2}\right)^{2}\,d^{2}n
  &=\frac{4}{15},\qquad\alpha\neq\beta,\\
  \frac{1}{4\pi}\oint_{\partial V}\left(n_{\alpha}^{2}-n_{\beta}^{2}\right)
  \left(n_{\alpha}^{2}-n_{\gamma}^{2}\right)\,d^{2}n
  &=\frac{2}{15},\qquad\alpha\neq\beta\neq\gamma,
  \label{eq:SomeIntegs}
\end{align}
are standard and can be evaluated analytically.

The main results are now summarized.

\subsubsection{First correction: $\overline{\boldmath{\psi}^{2}}$}

The averaged squared deviation is
\begin{align*}
  \overline{\bm{\psi}^{2}}
  &=\frac{1}{a_{0}^{3}}\int d\mathbf{r}\,\bm{\psi}^{2}
  =\frac{1}{a_{0}^{3}}\int d\mathbf{r}\,(\psi_{1}^{2}+\psi_{2}^{2})\\
  &=\frac{4}{15}\frac{\lambda_{s}^{2}}{1-m_{0,z}^{2}}
  \left[\left(1-\frac{\kappa_{1}^{2}}{14}\right)^{2}m_{0,x}^{2}m_{0,y}^{2}
  +\left(1-\frac{\kappa_{2}^{2}}{14}\right)^{2}m_{0,z}^{2}
  \left(m_{0,x}^{2}m_{0,y}^{2}+m_{0,x}^{4}+m_{0,y}^{4}\right)\right].
\end{align*}
This can be approximated to leading order in the $\kappa_{\beta}^{2}$ as
\begin{align*}
  \overline{\bm{\psi}^{2}}\simeq\frac{4\lambda_{s}^{2}}{15}
  \left[m_{0,x}^{2}m_{0,y}^{2}+m_{0,z}^{2}(m_{0,x}^{2}+m_{0,y}^{2})\right],
\end{align*}
with subleading corrections involving $\kappa_{\beta}^{2}$.

This term is of order $k_{s}^{2}$ and modifies the Zeeman contribution to the effective field via the factor $1-\tfrac{1}{2}\overline{\bm{\psi}^{2}}$ in Eq.~\eqref{eq:EM_m0_RHSinPsi}.

\subsubsection{Second correction: $\int\boldmath{\psi}^{2}\mathbf{H}_{A}$}

The term
\begin{align*}
  \int d\mathbf{r}'\,\bm{\psi}^{2}\mathbf{H}_{A}
\end{align*}
can be evaluated using the same surface integrals. 
Its contribution is of order $k_{s}^{3}$ and is therefore subdominant compared to the $k_{s}^{2}$ correction from $\overline{\bm{\psi}^{2}}$. 
In practice one may neglect this term at leading order, but a quantitative assessment would require a numerical evaluation for realistic parameter values.

\subsubsection{Third correction: $\int(\boldmath{\psi}\times\partial_{t}\boldmath{\psi})$}

The most interesting contribution comes from the term
\begin{align*}
  \int d\mathbf{r}'\bigl(\bm{\psi}\times\partial_{t}\bm{\psi}\bigr),
\end{align*}
which encodes dynamical corrections induced by the transverse modes. 
This term must be treated as a genuine vector product.  If the transverse field is written in a local moving basis,
\begin{align*}
  \bm{\psi}=\psi_{1}\mathbf{e}_{1}(\mathbf{m}_{0})+\psi_{2}\mathbf{e}_{2}(\mathbf{m}_{0}),\qquad
  \mathbf{e}_{1}\times\mathbf{e}_{2}=\mathbf{m}_{0},
\end{align*}
then the leading scalar amplitudes enter through the antisymmetric Berry combination
\begin{align*}
  \psi_{1}\dot{\psi}_{2}-\psi_{2}\dot{\psi}_{1},
\end{align*}
together with additional connection terms coming from the time dependence of the transverse basis $\mathbf{e}_{1,2}(\mathbf{m}_{0})$. Thus this contribution is not obtained from the symmetric combination
$\psi_{1}\dot{\psi}_{1}+\psi_{2}\dot{\psi}_{2}$, which would merely be proportional to $\partial_{t}(\psi_{1}^{2}+\psi_{2}^{2})$.
The detailed structure is summarized in Appendix~\ref{app:ThirdTerm}.
In compact form, one may write
\begin{align*}
  \int d\mathbf{r}'\bigl(\bm{\psi}\times\partial_{t}\bm{\psi}\bigr)
  =
  \int d\mathbf{r}'\left[
  (\psi_{1}\dot{\psi}_{2}-\psi_{2}\dot{\psi}_{1})\mathbf{m}_{0}
  +\bm{\mathcal{C}}(\psi_{1},\psi_{2};\mathbf{m}_{0},\dot{\mathbf{m}}_{0})
  \right],
\end{align*}
where $\bm{\mathcal{C}}$ denotes the moving-frame connection contribution. 
Both terms are of order $\lambda_{s}^{2}$ and linear in $\dot{\mathbf{m}}_{0}$ after insertion of the adiabatic solution for $\bm{\psi}$.  When inserted into Eq.~\eqref{eq:EM_m0_RHSinPsi}, they generate nonlinear dynamical corrections to the macrospin equation.

\subsection{Schematic form and interpretation of the correction terms}

The extended equation of motion for the macrospin $\mathbf{m}_{0}$ derived above can be written schematically as
\begin{align}
  \partial_{t}\mathbf{m}_{0}
  &=\mathbf{m}_{0}\times\mathbf{H}_{\mathrm{eff}}^{(0)}
  +\mathbf{m}_{0}\times\delta\mathbf{H}_{\mathrm{eff}}[\mathbf{m}_{0},\partial_{t}\mathbf{m}_{0}],
\end{align}
where $\mathbf{H}_{\mathrm{eff}}^{(0)}$ is the usual effective field (Zeeman plus anisotropy) of a macrospin and $\delta\mathbf{H}_{\mathrm{eff}}$ contains the surface-induced corrections.
The first two terms in $\delta\mathbf{H}_{\mathrm{eff}}$ [cf.\ Eq.~\eqref{eq:CorrectionField}] renormalize the precession (FMR) frequency by modifying the magnitude and direction of the effective field.
The third term, involving $\bm{\psi}\times\partial_{t}\bm{\psi}$, introduces a dependence on $\partial_{t}\mathbf{m}_{0}$ and thus acts as an effective damping/nutation contribution.
Importantly, this term is nonlocal in the magnetization components and contains high powers of $\mathbf{m}_{0}$, reflecting its origin in surface-induced spin misalignment.

\section{Discussion}
\label{sec:Discussion}

The LLE was previously extended beyond the macrospin approximation by accounting for surface-induced spin misalignment in a nanomagnet. This extension was achieved through the integration of transverse spin fluctuations (the fast degrees of freedom), resulting in a renormalized effective field within the LLE.
In the following, it is argued that this process and approach generalize well beyond the specific case of spin fluctuations caused by boundary effects.

\subsection{Practical relevance and generality}

Indeed, beyond the extension of the LLE derived above, two remarks are worth emphasizing regarding the practical relevance and generality of the present results.

This is a significant practical advantage of the present result: unlike the nutation resonance itself, which lies in the sub-THz/THz range and requires dedicated broadband or time-resolved setups, a shift of the FMR frequency and a change in its effective anisotropy field are directly accessible with standard, commercially available FMR spectrometers operating in the GHz range. In other words, the surface-induced correction captured by the first two terms of $\delta\mathbf{H}_{\mathrm{eff}}$ does not require resolving the nutation dynamics per se---its signature is already imprinted on the ordinary precessional response, through a measurable renormalization of the resonance field/frequency and linewidth. This low-frequency fingerprint of nutation was precisely the point made in Ref.~\citep{adamsLowfrequencySignatureMagnetization2024}, where the same class of surface corrections was shown to manifest as a detectable shift in the FMR spectrum rather than as a distinct high-frequency peak; the microscopic derivation presented here provides the effective-field expression [Eq.~\eqref{eq:CorrectionField}] that would enter such an analysis quantitatively. A similar coupling between inertial/nutational corrections and the ordinary FMR response (peak shift and broadening) was also reported phenomenologically in Ref.~\citep{oliveFerromagneticResonanceInertial2012}.

In fact, the mechanism responsible for the emergence of nutation in the present treatment is more general than the specific example considered here. Surface-anisotropy-induced spin misalignment, discussed above, is only one physical realization of a broader structural fact: whenever a slow, macroscopic dynamical variable is dynamically coupled to a bath of fast fluctuating degrees of freedom, adiabatic elimination of the fast variables generically dresses the slow variable's equation of motion with memory and inertia-like terms, of which nutation is the leading manifestation.
In the present case, the slow variable is the macrospin $\mathbf{m}_{0}$, and the ``bath'' is the transverse fluctuation field $\bm{\psi}$ generated by surface disorder; but the same structural argument applies whenever spin misalignment or disorder arises from other sources, e.g., lattice defects, temperature-driven fluctuations of the local magnetization~\citep{thibaudeauEmergingMagneticNutation2021}, non-adiabatic electron-spin coupling~\citep{fahnleGeneralizedGilbertEquation2011,kikuchiSpinDynamicsInertia2015}, or exchange with a nutation-carrying spin-wave continuum~\citep{makhfudzNutationWavePlatform2020}.
This places the present result within a well-known class of problems in nonequilibrium statistical mechanics, in which the reduction of a many-body/bath problem to a closed equation for a few slow variables is achieved through projection-operator or path-integral elimination techniques, and generically produces a memory kernel coupling the slow variable to its own past~\citep{moriTransportCollectiveMotion1965,zwanzigNonlinearGeneralizedLangevin1973,caldeiraPathIntegralApproach1983}.
In the Markovian (adiabatic) limit considered here, this memory kernel reduces to a term proportional to $\partial_{t}\mathbf{m}_{0}$, i.e., precisely a nutation/damping contribution; a non-adiabatic treatment retaining the full time dependence of $\bm{\psi}$ would instead yield a genuinely retarded (non-Markovian) equation of motion for $\mathbf{m}_{0}$, of which the present result is the leading-order truncation.
As a general statement, this suggest that nutational dynamics should be expected as a generic companion of magnetization dynamics whenever the macrospin is coupled, through any microscopic mechanism, to a reservoir of fast transverse fluctuations, and that the present formalism offers a systematic route to compute the corresponding effective field for a given microscopic source of disorder.

\subsection{Comparison with existing approaches}

It is instructive to compare this structure with inertial extensions of the Landau--Lifshitz--Gilbert equation~\citep{ciorneiMagnetizationDynamicsInertial2011,oliveFerromagneticResonanceInertial2012,oliveDeviationLandauLifshitzGilbertEquation2015,wegroweMagneticMonopoleSeparation2016}, where a second-order time derivative of the magnetization is added phenomenologically, leading to nutation on top of precession.
In the present microscopic approach, the adiabatic elimination of fast transverse modes generates an effective first-order equation in which the extra term arises from the Berry phase coupling between $\mathbf{m}_{0}$ and $\bm{\psi}$.
Formally, this term can be interpreted as a renormalization of the Berry phase coefficient and effective damping kernel, and a more systematic treatment (beyond the adiabatic approximation) would naturally lead to memory effects and possibly higher-order derivatives.

It is noted that both the present approach and the effective one-spin particle (EOSP) method employed in Ref.~\citep{adamsSpatialMagnetizationProfile2023} lead to modifications of the effective field in the Landau--Lifshitz equation. However, a distinct difference is observed in the explicit form of these corrections: the dominant contribution scales as $m_{0}^{2}$ in the adiabatic approach, whereas a scaling of $m_{0}^{3}$ is obtained in the EOSP framework. A quantitative comparison of these corrections, including estimates of the induced nutation frequency and damping for realistic nanoparticle parameters, is deferred to future investigations.

Several directions for future study are suggested by this work. On the theoretical side, an explicit memory kernel for the macrospin dynamics could be derived by extending the analysis beyond the current adiabatic (Markovian) approximation, specifically by retaining the full time dependence of the transverse fluctuations rather than utilizing their adiabatically eliminated solution. Such an extension would provide a quantitative test of the proposed generic bath-induced nutation mechanism and clarify its range of validity. On the experimental side, a concrete, readily testable prediction for nanomagnets with well-characterized surface anisotropy is constituted by the predicted renormalization of the FMR frequency and lineshape [Eq.~\eqref{eq:CorrectionField}]. This complements the low-frequency nutation signatures already reported in Ref.~\citep{adamsLowfrequencySignatureMagnetization2024}, and it is anticipated that a systematic study of its dependence on particle size, anisotropy strength, and temperature would facilitate its discrimination from alternative sources of FMR linewidth broadening.

Finally, applying the present formalism to other realizations of the slow/fast variable separation discussed above, for instance nanomagnets coupled to a magnon bath, antiferromagnetic or synthetic-ferrimagnetic structures, or spin-torque nano-oscillators, would test the generality of the proposed nutation-as-dressing mechanism and help delineate which physical systems should be expected to exhibit an observable nutation signal.

\newpage
\appendix
%% Magnetization nutation in nanomagnets: extended LLE in the adiabatic approximation
%% Optimized draft aimed at a JMP / theory-oriented audience.

\section{Spin Berry phase and curvature}
\label{app:Spin_Berry_phase_and_curvature}

\newcommand{\dd}{\mathrm d}
\newcommand{\ii}{\mathrm i}
\newcommand{\m}{\mathbf m}
\newcommand{\Acal}{\mathcal A}
\newcommand{\Scal}{\mathcal S}
This appendix provides explicit forms of the spin Berry phase and curvature for spin-coherent states.

Starting from the Euclidean coherent-state action,
\begin{align}
  \Scal_{\mathrm E}[\m]
  &=
  \int_0^{\beta\hbar} \dd\tau\,
  \Braket{\m(\tau)|\frac{\dd}{\dd\tau}\m(\tau)}
  +\frac{1}{\hbar}
  \int_0^{\beta\hbar}\dd\tau\,
  \Braket{\m(\tau)|\mathcal H|\m(\tau)},
  \label{eq:SE_start}
\end{align}
the first term is isolated,
\begin{align}
  \Scal_{\mathrm B}[\m]
  &\equiv
  \int_0^{\beta\hbar}\dd\tau\,
  \Braket{\m(\tau)|\frac{\dd}{\dd\tau}\m(\tau)} .
  \label{eq:SB_def}
\end{align}
This is the Berry contribution to the Euclidean action. Because the spin coherent states are normalized,
\begin{align}
  \Braket{\m|\m}=1,
\end{align}
one has
\begin{align}
  \dd\Braket{\m|\m}
  &=
  \Braket{\dd\m|\m}
  +
  \Braket{\m|\dd\m}=0,
\end{align}
and therefore
\begin{align}
  \Braket{\m|\dd\m}^{*}
  &=
  \Braket{\dd\m|\m}
  =
  -\Braket{\m|\dd\m}.
\end{align}
Thus the Berry one-form
\begin{align}
  \Acal(\m)
  \equiv
  \Braket{\m|\dd\m}
  \label{eq:Berry_one_form}
\end{align}
is purely imaginary.

The Berry term in Eq.~\eqref{eq:SE_start} is the line integral of this one-form along the closed trajectory
\(C:\tau\mapsto\m(\tau)\) on the unit sphere,
\begin{align}
  \Scal_{\mathrm B}[\m]
  =
  \oint_C \Acal .
  \label{eq:Berry_line_integral}
\end{align}
The trajectory is closed because of the periodic boundary condition in imaginary time,
\(\m(\beta\hbar)=\m(0)\), up to the usual coherent-state phase convention.

To convert this line integral into a surface integral, a smooth extension is introduced,
\begin{align}
  \m(\tau,u),
  \qquad
  0\leq u\leq 1,
\end{align}
with
\begin{align}
  \m(\tau,1)=\m(\tau),
  \qquad
  \m(\tau,0)=\m_{\rm ref},
  \qquad
  \m(0,u)=\m(\beta\hbar,u),
  \label{eq:extension_conditions}
\end{align}
where \(\m_{\rm ref}\) is an arbitrary fixed reference spin direction. This extension fills a surface \(\Sigma\) on the unit sphere whose boundary is the physical trajectory \(C\). Stokes' theorem gives
\begin{align}
  \oint_C \Acal
  =
  \int_{\Sigma} \dd\Acal .
  \label{eq:Stokes}
\end{align}
Using \(\Acal=\Braket{\m|\dd\m}\), its exterior derivative is
\begin{align}
  \dd\Acal
  &=
  \dd\Braket{\m|\dd\m}
  =
  \Braket{\dd\m|\wedge|\dd\m} .
  \label{eq:curvature_general}
\end{align}
Here \(\Braket{\dd\m|\wedge|\dd\m}\) denotes the Berry-curvature two-form on the coherent-state manifold. Pulling this two-form back to the \((\tau,u)\) surface yields
\begin{align}
  \Scal_{\mathrm B}[\m]
  &=
  \int_0^{\beta\hbar}\dd\tau
  \int_0^1\dd u\,
  \Braket{
  \partial_\tau\m(\tau,u)
  \left|\wedge\right|
  \partial_u\m(\tau,u)} .
  \label{eq:eq10_intermediate}
\end{align}
This is the differential-geometric content of Eq.~(10): the Berry phase line integral is rewritten as the flux of the Berry curvature through a surface bounded by the spin trajectory.

For spin coherent states of spin length \(S\), the Berry curvature on the sphere is proportional to the area form.  With the sign convention adopted here, one may write
\begin{align}
  \Braket{\dd\m|\wedge|\dd\m}
  =
  -\ii S\,
  \m\cdot(\dd\m\times\dd\m),
  \label{eq:curvature_sphere_compact}
\end{align}
or, equivalently, after pullback to the \((\tau,u)\) surface,
\begin{align}
  \Braket{
  \partial_\tau\m
  \left|\wedge\right|
  \partial_u\m}
  =
  -\ii S\,
  \m(\tau,u)\cdot
  \left(
  \partial_\tau\m(\tau,u)
  \times
  \partial_u\m(\tau,u)
  \right).
  \label{eq:curvature_pulled_back}
\end{align}
Therefore
\begin{align}
  \int_0^{\beta\hbar}\dd\tau\,
  \Braket{\m(\tau)|\frac{\dd}{\dd\tau}\m(\tau)}
  &=-\ii S
  \int_0^{\beta\hbar}\dd\tau
  \int_0^1\dd u\,
  \m(\tau,u)\cdot
  \left(
  \partial_\tau\m(\tau,u)
  \times
  \partial_u\m(\tau,u)
  \right) .
  \label{eq:Berry_to_WZW}
\end{align}
This identifies the real Wess--Zumino--Witten functional as
\begin{align}
  \Scal_{\mathrm{WZW}}[\m]
  &\equiv
  \int_0^{\beta\hbar}\dd\tau
  \int_0^1\dd u\,
  \m(\tau,u)\cdot
  \left(
  \partial_\tau\m(\tau,u)
  \times
  \partial_u\m(\tau,u)
  \right),
  \label{eq:WZW_def_final}
\end{align}
so that the Berry part of the Euclidean action is
\begin{align}
  \Scal_{\mathrm B}[\m]
  =
  -\ii S\,\Scal_{\mathrm{WZW}}[\m].
  \label{eq:Berry_equals_minus_i_WZW}
\end{align}
Consequently, the full Euclidean action becomes
\begin{align}
  \Scal_{\mathrm E}[\m]
  =
  -\ii S\,\Scal_{\mathrm{WZW}}[\m]
  +
  \frac{1}{\hbar}
  \int_0^{\beta\hbar}\dd\tau\,
  \Braket{\m(\tau)|\mathcal H|\m(\tau)} .
  \label{eq:SE_final}
\end{align}
This is the form used later in the main text.

\paragraph*{Gauge and orientation convention.}
The overall sign in Eqs.~\eqref{eq:curvature_pulled_back}--\eqref{eq:Berry_equals_minus_i_WZW} depends on the orientation chosen for the auxiliary surface and on the coherent-state gauge.  Reversing the order of the cross product, \(\partial_\tau\m\times\partial_u\m\to\partial_u\m\times\partial_\tau\m\), changes the sign.  Once a convention is chosen, the same convention must be used consistently in the Euclidean action, in the real-time magnetic action, and in the variation of the WZW term.

\paragraph*{Coordinate check.}
For completeness, take the usual coherent-state gauge in which
\begin{align}
  \Braket{\m|\dd\m}
  &=
  -\ii S(1-\cos\theta)\,\dd\varphi .
  \label{eq:gauge_choice}
\end{align}
Then
\begin{align}
  \dd\Braket{\m|\dd\m}
  &=
  -\ii S\sin\theta\,\dd\theta\wedge\dd\varphi .
\end{align}
On the other hand,
\begin{align}
  \m\cdot(\partial_\tau\m\times\partial_u\m)
  &=
  \sin\theta
  \left(
  \partial_\tau\theta\,\partial_u\varphi
  -
  \partial_u\theta\,\partial_\tau\varphi
  \right),
\end{align}
which confirms Eq.~\eqref{eq:curvature_pulled_back} with the orientation used above.

\section{Landau--Lifshitz equation for a ferromagnet}
\label{app:LLE}

For completeness, the derivation of the classical Landau--Lifshitz equation is recalled from the continuum action~\eqref{eq:WZW_MSP_Continuum}, following the approach established by Fradkin~\citep{fradkinFieldTheoriesCondensed2015}. The constraint $\mathbf{m}^{2}=1$ is enforced by the introduction of a Lagrange multiplier $\lambda(\mathbf{r},t)$ and the inclusion of the term
\begin{align}
  \mathcal{S}_{\mathrm{L}}[\lambda,\mathbf{m}]
  &=\frac{1}{2}\int_{0}^{T}\!dt\int d\mathbf{r}\,
  \lambda(\mathbf{r},t)\left[\mathbf{m}^{2}(\mathbf{r},t)-1\right].
  \label{eq:ConstraintAction-app}
\end{align}
The total action is
\begin{align*}
  \mathcal{S}_{\mathrm{T}}[\mathbf{m}]
  &=\mathcal{S}_{\mathrm{M}}[\mathbf{m}]+\mathcal{S}_{\mathrm{L}}[\lambda,\mathbf{m}],
\end{align*}
with $\mathcal{S}_{\mathrm{M}}$ given by Eq.~\eqref{eq:WZW_MSP_Continuum}.

The variation of the WZW term with respect to $\mathbf{m}$ is
\begin{align}
  \delta\mathcal{S}_{\mathrm{WZ}}&=
  \int_{0}^{T}\!dt\int d\mathbf{r}\delta\mathbf{m}\cdot
  \left(\mathbf{m}\times\partial_{t}\mathbf{m}\right).
  \label{eq:Action_WZW_var}
\end{align}
The Euler--Lagrange equations then read
\begin{align*}
  \frac{\delta\mathcal{S}_{\mathrm{T}}}{\delta\mathbf{m}}
  -\sum_{\mu}\partial^{\mu}\left(
  \frac{\delta\mathcal{S}_{\mathrm{T}}}{\delta(\partial^{\mu}\mathbf{m})}
  \right)&=0,\qquad\mu=0,1,2,3.
\end{align*}
Carrying out the variations, one finds
\begin{align}
  \frac{S}{a_{0}^{3}}\left(\mathbf{m}\times\partial_{t}\mathbf{m}\right)
  +\frac{\mu_{a}S}{a_{0}^{3}}\mathbf{H}
  +\frac{S^{2}}{a_{0}^{3}}\frac{\delta\mathcal{A}[\mathbf{m}]}{\delta\mathbf{m}}
  +\lambda\mathbf{m}
  &=-\frac{S^{2}}{a_{0}}J(\mathbf{r},\mathbf{r}')\,\Delta\mathbf{m}(\mathbf{r},t).
  \label{eq:n-EM}
\end{align}
The functional derivative of the anisotropy functional is
\begin{align*}
  \frac{\delta\mathcal{A}[\mathbf{m}]}{\delta\mathbf{m}}=
  \begin{cases}
    2K_{c}(\mathbf{m}\cdot\mathbf{e}_{z})\mathbf{e}_{z}, & \mathbf{r}\in\mathrm{core},\\[3pt]
    -K_{s}\displaystyle\sum_{\boldsymbol{\delta}}(\mathbf{m}\cdot\boldsymbol{\delta})\boldsymbol{\delta},
    & \mathbf{r}\in\mathrm{surface}.
  \end{cases}
\end{align*}
The Lagrange multiplier $\lambda$ can be found by taking the scalar product of Eq.~\eqref{eq:n-EM} with $\mathbf{m}$. After some algebra, one obtains the equation
\begin{align*}
  \mathbf{m}\times\partial_{t}\mathbf{m}
  +\mathbf{m}\times\left(\mathbf{H}_{\mathrm{eff}}\times\mathbf{m}\right)&=0,
\end{align*}
with effective field
\begin{align}
  \mathbf{H}_{\mathrm{eff}}&=Sa_{0}^{2}J\,\Delta\mathbf{m}
  +\mu_{a}\mathbf{H}+\mathbf{H}_{A},
  \label{eq:n-EF}
\end{align}
where $\mathbf{H}_{A}$ is the anisotropy field. By multiplying by $\mathbf{m}$ and employing the constraint $\mathbf{m}^{2}=1$, the undamped Landau--Lifshitz equation is obtained as
\begin{align}
  \partial_{t}\mathbf{m}&=\mathbf{m}\times\mathbf{H}_{\mathrm{eff}}
  =\mathbf{m}\times\left[Sa_{0}^{2}J\,\Delta\mathbf{m}
  +\mu_{a}\mathbf{H}+\mathbf{H}_{A}\right].
  \label{eq:n-LLE}
\end{align}

\section{Evaluation of the third term}
\label{app:ThirdTerm}

In this appendix, the explicit evaluation of the physical-boundary WZW contribution retained in Eq.~\eqref{eq:Action_S2} is provided.
\begin{align}
  \mathbf{I}_{\psi}
  \equiv
  \int d\mathbf{r}'\bigl(\bm{\psi}\times\partial_{t}\bm{\psi}\bigr),
  \label{eq:Ipsi_def}
\end{align}
appears in the correction field in Eq.~\eqref{eq:CorrectionField}.  Since this contribution comes from the WZW part of the real-time magnetic action $\mathcal{S}_{\mathrm{M}}$, it must retain the antisymmetric Berry structure. Therefore, if
\begin{align}
  \bm{\psi}
  =\psi_{1}\mathbf{e}_{1}(t)+\psi_{2}\mathbf{e}_{2}(t),
  \qquad
  \mathbf{e}_{1}\times\mathbf{e}_{2}=\mathbf{m}_{0},
  \label{eq:psi_moving_basis_app}
\end{align}
where $\mathbf{e}_{1,2}$ span the plane transverse to $\mathbf{m}_{0}$, one has
\begin{align}
  \begin{aligned}
  \bm{\psi}\times\partial_{t}\bm{\psi}
  =&\,
  \bigl(\psi_{1}\dot{\psi}_{2}-\psi_{2}\dot{\psi}_{1}\bigr)\mathbf{m}_{0}
  +\psi_{1}^{2}\,\mathbf{e}_{1}\times\dot{\mathbf{e}}_{1}
  +\psi_{2}^{2}\,\mathbf{e}_{2}\times\dot{\mathbf{e}}_{2}
  \\
  &+\psi_{1}\psi_{2}
  \left(
  \mathbf{e}_{1}\times\dot{\mathbf{e}}_{2}
  +\mathbf{e}_{2}\times\dot{\mathbf{e}}_{1}
  \right).
  \end{aligned}
  \label{eq:psi_cross_dpsi_full_app}
\end{align}
Thus the amplitude part is governed by the antisymmetric combination
$\psi_{1}\dot{\psi}_{2}-\psi_{2}\dot{\psi}_{1}$, not by the symmetric expression
$\psi_{1}\dot{\psi}_{1}+\psi_{2}\dot{\psi}_{2}$. The latter is only
$\frac{1}{2}\partial_{t}(\psi_{1}^{2}+\psi_{2}^{2})$ and does not represent the Berry/WZW cross product.

By keeping the explicit surface-induced amplitudes presented in Eq.~\eqref{eq:psi-LargestDev}, and assuming the time derivatives of the coefficients $\kappa_{1,2}$ are negligible at this order of approximation, the following definitions are established:
\begin{align*}
  c_{1}=1-\frac{\kappa_{1}^{2}}{14},
  \qquad
  c_{2}=1-\frac{\kappa_{2}^{2}}{14},
  \qquad
  q=1-m_{0,z}^{2}=m_{0,x}^{2}+m_{0,y}^{2}.
\end{align*}
The two mixed terms entering the antisymmetric Berry combination are then
\begin{align}
  \begin{aligned}
  \int d\mathbf{r}'\,\psi_{1}\partial_{t}\psi_{2}
  =&\frac{2\lambda_{s}^{2}}{15}c_{1}c_{2}
  \left[
  \frac{2m_{0,x}^{2}m_{0,y}m_{0,z}}{q}\,\dot{m}_{0,x}
  -\frac{2m_{0,x}m_{0,y}^{2}m_{0,z}}{q}\,\dot{m}_{0,y}
  \right.
  \\
  &\left.
  +\frac{m_{0,x}m_{0,y}(m_{0,x}^{2}-m_{0,y}^{2})}{q^{2}}\,\dot{m}_{0,z}
  \right],
  \end{aligned}
  \label{eq:psi1dpsi2}
\end{align}
and
\begin{align}
  \begin{aligned}
  \int d\mathbf{r}'\,\psi_{2}\partial_{t}\psi_{1}
  =&\frac{2\lambda_{s}^{2}}{15}c_{1}c_{2}
  \left[
  \frac{m_{0,y}m_{0,z}(m_{0,x}^{2}-m_{0,y}^{2})}{q}\,\dot{m}_{0,x}
  +\frac{m_{0,x}m_{0,z}(m_{0,x}^{2}-m_{0,y}^{2})}{q}\,\dot{m}_{0,y}
  \right.
  \\
  &\left.
  +\frac{m_{0,x}m_{0,y}m_{0,z}^{2}(m_{0,x}^{2}-m_{0,y}^{2})}{q^{2}}\,\dot{m}_{0,z}
  \right].
  \end{aligned}
  \label{eq:psi2dpsi1}
\end{align}
Subtracting Eq.~\eqref{eq:psi2dpsi1} from Eq.~\eqref{eq:psi1dpsi2} gives the amplitude part of the WZW correction,
\begin{align}
  \begin{aligned}
  \mathcal{I}_{\mathrm{amp}}
  &\equiv
  \int d\mathbf{r}'\,
  \bigl(\psi_{1}\partial_{t}\psi_{2}-\psi_{2}\partial_{t}\psi_{1}\bigr)
  \\
  &=\frac{2\lambda_{s}^{2}}{15}c_{1}c_{2}
  \left[
  m_{0,z}m_{0,y}\,\dot{m}_{0,x}
  -m_{0,z}m_{0,x}\,\dot{m}_{0,y}
  +\frac{m_{0,x}m_{0,y}(m_{0,x}^{2}-m_{0,y}^{2})}{q}\,\dot{m}_{0,z}
  \right].
  \end{aligned}
  \label{eq:Iamp_final}
\end{align}
Consequently, the part of $\mathbf{I}_{\psi}$ associated with the antisymmetric scalar amplitudes is
\begin{align}
  \mathbf{I}_{\psi}^{\mathrm{amp}}
  =\mathcal{I}_{\mathrm{amp}}\,\mathbf{m}_{0}.
  \label{eq:Ipsi_amp_final}
\end{align}
This is the direct correction to the expression that previously involved the symmetric combination.

The remaining terms in Eq.~\eqref{eq:psi_cross_dpsi_full_app} come from the time dependence of the transverse frame.  In a parallel-transported frame, chosen such that the residual rotation around $\mathbf{m}_{0}$ is zero,
\begin{align}
  \dot{\mathbf{e}}_{a}=-(\mathbf{e}_{a}\cdot\dot{\mathbf{m}}_{0})\mathbf{m}_{0},
  \qquad a=1,2,
  \label{eq:parallel_frame_app}
\end{align}
and Eq.~\eqref{eq:psi_cross_dpsi_full_app} becomes
\begin{align}
  \begin{aligned}
  \bm{\psi}\times\partial_{t}\bm{\psi}
  =&\,
  \bigl(\psi_{1}\dot{\psi}_{2}-\psi_{2}\dot{\psi}_{1}\bigr)\mathbf{m}_{0}
  +\left[\psi_{2}^{2}(\mathbf{e}_{2}\cdot\dot{\mathbf{m}}_{0})
  +\psi_{1}\psi_{2}(\mathbf{e}_{1}\cdot\dot{\mathbf{m}}_{0})\right]\mathbf{e}_{1}
  \\
  &-
  \left[\psi_{1}^{2}(\mathbf{e}_{1}\cdot\dot{\mathbf{m}}_{0})
  +\psi_{1}\psi_{2}(\mathbf{e}_{2}\cdot\dot{\mathbf{m}}_{0})\right]\mathbf{e}_{2}.
  \end{aligned}
  \label{eq:psi_cross_dpsi_parallel_app}
\end{align}
For completeness, the required surface averages appearing in the connection part are
\begin{align}
  \int d\mathbf{r}'\,\psi_{1}^{2}
  &=\frac{4\lambda_{s}^{2}}{15}c_{1}^{2}\frac{m_{0,x}^{2}m_{0,y}^{2}}{q},
  \label{eq:int_psi1sq_app}
  \\
  \int d\mathbf{r}'\,\psi_{2}^{2}
  &=\frac{4\lambda_{s}^{2}}{15}c_{2}^{2}\frac{m_{0,z}^{2}}{q}
  \left(m_{0,x}^{2}m_{0,y}^{2}+m_{0,x}^{4}+m_{0,y}^{4}\right),
  \label{eq:int_psi2sq_app}
  \\
  \int d\mathbf{r}'\,\psi_{1}\psi_{2}
  &=\frac{2\lambda_{s}^{2}}{15}c_{1}c_{2}\frac{m_{0,x}m_{0,y}m_{0,z}}{q}
  \left(m_{0,x}^{2}-m_{0,y}^{2}\right).
  \label{eq:int_psi1psi2_app}
\end{align}
Combining Eqs.~\eqref{eq:Ipsi_amp_final} and~\eqref{eq:psi_cross_dpsi_parallel_app} gives the complete WZW-induced correction in the chosen transverse-frame gauge,
\begin{align}
  \begin{aligned}
  \mathbf{I}_{\psi}
  =&\,\mathcal{I}_{\mathrm{amp}}\,\mathbf{m}_{0}
  +\left[
  \left(\int d\mathbf{r}'\,\psi_{2}^{2}\right)(\mathbf{e}_{2}\cdot\dot{\mathbf{m}}_{0})
  +\left(\int d\mathbf{r}'\,\psi_{1}\psi_{2}\right)(\mathbf{e}_{1}\cdot\dot{\mathbf{m}}_{0})
  \right]\mathbf{e}_{1}
  \\
  &-
  \left[
  \left(\int d\mathbf{r}'\,\psi_{1}^{2}\right)(\mathbf{e}_{1}\cdot\dot{\mathbf{m}}_{0})
  +\left(\int d\mathbf{r}'\,\psi_{1}\psi_{2}\right)(\mathbf{e}_{2}\cdot\dot{\mathbf{m}}_{0})
  \right]\mathbf{e}_{2}.
  \end{aligned}
  \label{eq:Ipsi_complete_app}
\end{align}
This expression is of order $\lambda_{s}^{2}$ and linear in $\dot{\mathbf{m}}_{0}$.  When inserted into Eq.~\eqref{eq:CorrectionField}, it provides the dynamical surface-induced correction to the effective field.  The important point is that the sign and structure are fixed by the antisymmetric Berry combination in Eq.~\eqref{eq:Iamp_final}; using $\psi_{1}\dot{\psi}_{1}+\psi_{2}\dot{\psi}_{2}$ would instead produce a total time derivative of $\bm{\psi}^{2}$ and would not yield the WZW torque.

\bibliographystyle{apsrev4-2}
%\bibliography{HamidKlib}
%

\end{document}